\documentclass[useAMS,usenatbib]{mn2e}

\usepackage{graphicx}
\usepackage{graphics}
\usepackage{natbib}
\usepackage{amsmath}
\usepackage{subfig}
\usepackage{color}
\usepackage{amssymb}

\usepackage{gensymb}

\title[Crab Optical and $\gamma$-Ray polarization]{A recent change in the optical and $\gamma$-ray polarization of the Crab nebula and pulsar}

\author[P. Moran, G. Kyne, C. Gouiff\`es, P. Laurent, G. Hallinan, R. M. Redfern, A. Shearer]{P. Moran$^{1}$,  G. Kyne$^{1,5}$, C. Gouiff\`es$^{2,3}$, P. Laurent$^{2,4}$, G. Hallinan$^{5}$, R. M. Redfern$^{1}$,
  \newauthor  A. Shearer$^{1}$\\
$^{1}$ Centre for Astronomy, School of Physics, National University of Ireland Galway, Ireland\\
$^{2}$ DSM/Irfu/Service d'Astrophysique, Bat. 709 Orme des Merisiers CEA Saclay, 91191 Gif-sur-Yvette Cedex, France\\
$^{3}$ Laboratoire AIM, UMR 7158 (CEA/Irfu, CNRS/INSU, Universit\'e Paris VII), CEA Saclay, B�t. 709, \\
 F-91191 Gif-sur-Yvette Cedex, France\\
$^{4}$ AstroParticule et Cosmologie, UMR 7164 (Universit\'e Paris Diderot, CNRS/IN2P3, CEA/DSM, Observatoire de Paris, \\  
Sorbonne Paris Cit\'e) 10, rue Alice Domon et Lonie Duquet, F-75205 Paris Cedex 13, France\\
$^{5}$ Department of Astronomy, Caltech MC 249-17, 1200 East California Blvd, Pasadena CA 91125}

\begin{document}

\date{}

\pagerange{\pageref{firstpage}--\pageref{lastpage}} \pubyear{2002}

\maketitle

\label{firstpage}

\begin{abstract}
We report on observations of the polarization of optical and $\gamma$-ray photons from the Crab nebula and pulsar system using the Galway Astronomical Stokes Polarimeter (GASP), the Hubble Space Telescope/Advanced Camera for Surveys (HST/ACS) and the International Gamma-Ray Astrophysics Laboratory satellite (Integral). These, when combined with other optical polarization observations, suggest that the polarized optical emission  and $\gamma$-ray polarization changes in a similar manner. A change in the optical polarization angle has been observed by this work, from $109.5 \pm0.7\degree$ in 2005 to  $85.3\pm1.4\degree$ in 2012. On the other hand, the $\gamma$-ray polarization angle changed from $115\pm11\degree$ ~in 2003--2007 to $80\pm12\degree$ ~in 2012--2014. Strong flaring activities have been detected in the Crab nebula over the past few years by the high energy $\gamma$-ray missions Agile and Fermi, and magnetic reconnection processes have been suggested to explain these observations. The change in the polarized optical and $\gamma$-ray emission of the Crab nebula/pulsar as observed, for the first time, by GASP and Integral may indicate that reconnection is possibly at work in the Crab nebula. %{\ Follow up GASP observations are planned for December 2015 to confirm this detected change in angle}. 
We also report, for the first time, a non-zero measure of the optical circular polarization from the Crab pulsar+knot system.

\end{abstract}

\begin{keywords}
polarization -- radiation mechanisms: non-thermal -- stars: neutron -- pulsars: individual: the Crab pulsar.
\end{keywords}

\section{Introduction}

The Crab pulsar, located in the Crab nebula, is a remnant of the supernova which occurred in 1054 \citep{Duyvenday42,Mayall42}. As such the pulsar is one of the rare examples of a pulsar of known age. As a young pulsar it powers its surrounding nebula that radiates at all electromagnetic frequencies from the radio to TeV $\gamma$-rays \citep{Hester08}. Polarized emission from the nebula and pulsar have been described by a number of authors; optical  \citep{Smith88, Slowikowska09, Moran13}; X-ray \citep{Weisskopf78} and $\gamma$-ray \citep{Dean08,Forot08} with the IBIS and SPI telescopes onboard Integral \citep{Ubertini03,Vedrenne03}. In particular the phase-resolved optical observations show a change in polarization consistent in shape with a beam of synchrotron radiation being beamed from both poles of an orthogonal rotator. Furthermore, \citet{Slowikowska09} indicate the presence of a highly polarized DC component which most likely corresponds to the nearby synchrotron emitting knot \citep{Moran13}.  Since the original Integral $\gamma$-ray observations, the Crab has been observed twice a year improving the statistics for determining the $\gamma$-ray polarization. 

Optical and X-ray observations show spatial and some flux variability of the inner nebula \citep{Bietenholz01}. Indeed this was observed in some early optical studies \citep{Scargle69}. However, the flux from the whole nebula was expected to be constant at the level of a few percent \citep{Kirsch05,Weisskopf10}. Since 2007 strong $\gamma$-ray flares have been observed by the Agile and Fermi $\gamma$-ray telescopes \citep{Abdo11,Tavani11,Striani13} at a rate of about 1 per year.  Subsequent X-ray, optical and radio observations did not show any significant flux variation across the nebula \citep{Lobanov11, Morii11ab, Weisskopf13} during these flaring events. The flares are of relatively short duration lasting from about 4 to 16 days. Within the flares, no changes of the pulsar's temporal properties such as spin-down rate were observed \citep{Abdo11}.  Here we report upon polarization studies of the Crab pulsar at optical and $\gamma$-ray wavelengths. Our aim was to see if the polarization was different after the flare events of 2010 to 2013 and whether the optical and $\gamma$-ray polarization was similar, possibly indicating a common origin. Throughout this paper the term polarization refers to linear polarization unless otherwise explicitly specified.

\section{Observations}

\begin{figure}
\centering
\includegraphics[width=90mm]{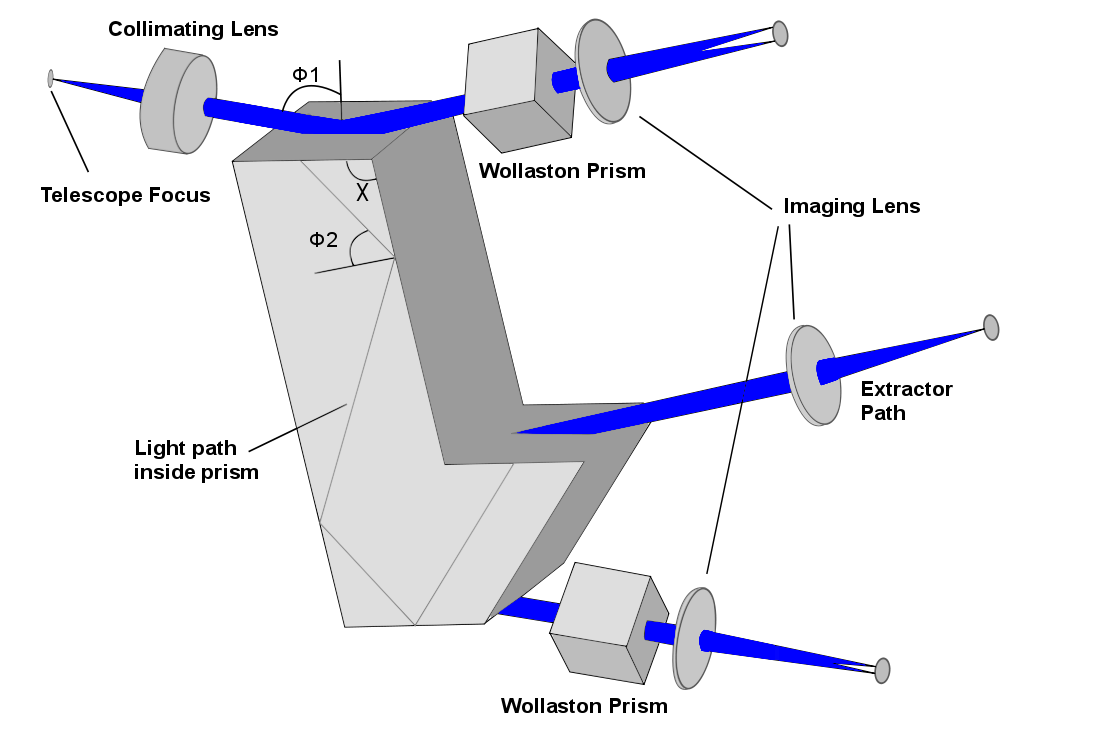}
\caption{2D optical layout of the GASP instrument using 2 Andor iXon Ultra 897 EMCCDs. The beam is split by the main beamsplitter prism Fresnel-rhomb generating the reflected and transmitted paths, RP and TP respectively. Each of these beams is passed through a Wollaston prism (polarizing beamsplitter) resulting in 4 intensity beams. A 90\degree ~phase shift is generated by the TP (which essentially acts as a quarter-wave plate) which measures circular polarization. The two images from the reflected path are shown in Figure 2.}
\label{Figure1}
\end{figure}

\begin{figure}
\centering
\includegraphics[height=50mm]{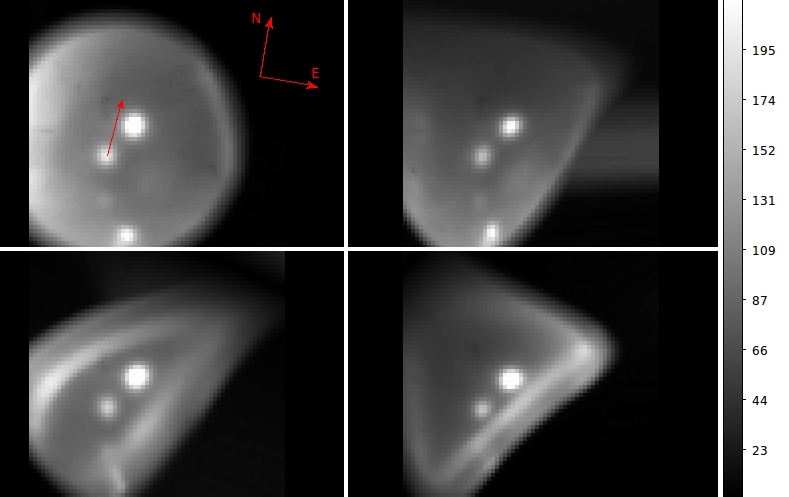}
\caption{The reflected and transmitted path images, RP1 (top left), RP2 (top right), TP1 (bottom left) and TP2 (bottom right), of a small section of the Crab nebula ($\sim$ 20\arcsec) where the Crab pulsar and the star Trimble 28, located North-East of the pulsar \citep{Trimble68}, can be seen. The polarization angle from the pulsar/knot is marked. It is clear that GASP suffers from large optical distortion errors, in particular the TP images; it was not possible to completely remove this using image registration techniques during post-processing.}
\label{Figure2}
\end{figure}

\begin{figure}
\centering
\includegraphics[width=80mm]{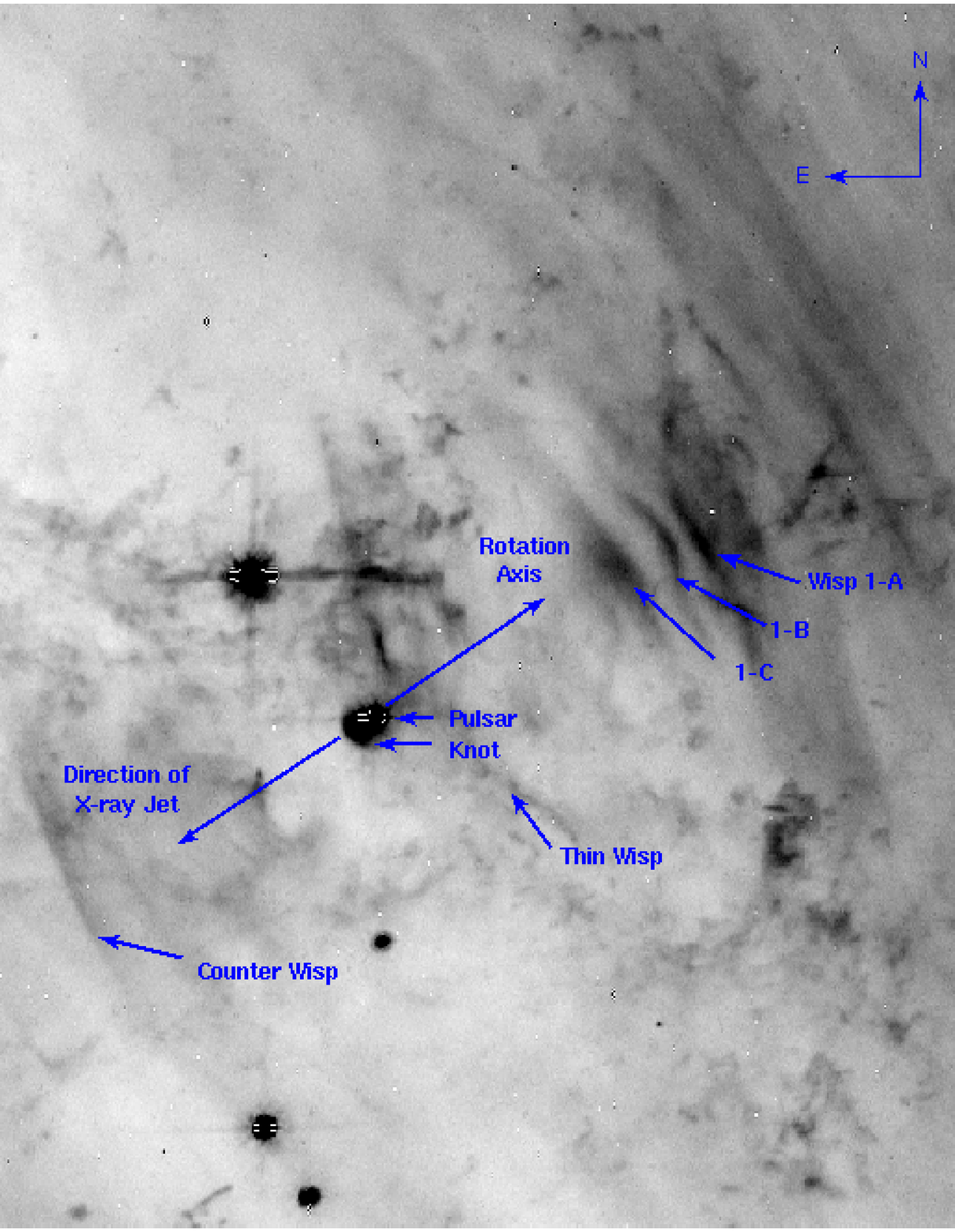}
\caption{Wide field image of the Crab nebula \citep{Moran13} showing the location of the knot with respect to the other features in the inner nebula. }
\label{Figure3}
\end{figure}

Optical polarimetry observations of the Crab pulsar were taken on 2012 November 12 using GASP \citep{Collins13} on the 200$^{\prime\prime}$ Hale telescope. Although GASP is normally capable of faster observations these were time averaged with an exposure time of 3 minutes using an R band filter centred at 650 nm. GASP used 2 Andor iXon Ultra 897 EMCCDs for these observations, and for this particular data set used a low level EM gain ($\sim$ 10) to improve the signal-to-noise ratio. These detectors were liquid cooled using an optimum temperature setting of -90\degree C. A set of polarization standards were also observed on the same night during suitable seeing, but variable atmospheric, conditions. Figure \ref{Figure1} shows the optical layout of GASP and Figure \ref{Figure2} shows the reflected and transmitted (RP and TP) images of the nebula. These images are subject to optical distortion, which has been measured and discussed in greater detail by \cite{Kyne14}. It was not possible to completely remove this in post-processing. As a comparison to the GASP measurement of the pulsar+knot polarization and to determine for comparison the polarization pre 2012, we re-examined the HST/ACS observations of the Crab from 2005 (see Figure \ref{Figure3}). 

The Crab nebula is repeatedly observed by Integral/IBIS \citep{Ubertini03} since its launch in 2002. The angular resolution of IBIS is $\sim$ 12\arcmin, hence the IBIS measurement encompasses the entire nebula and pulsar. In order to compare with the GASP observations only phase-averaged polarization is considered in the 300--450 keV energy band, which was chosen to optimize the signal to noise ratio.

\section{Data Reduction}

\subsection{Optical Data}
As an imaging polarimeter GASP is calibrated using a pixel-by-pixel Eigenvalue Calibration Method (ECM) \citep{Compain99}, to establish a uniform calibration. Errors in focus, image registration, and optical distortion can contribute to a non-uniform field of view (FOV) calibration; if this is the case, then it will not be possible to spatially recover the target's Stokes parameters correctly, or in any case it will limit the FOV over which the calibration is valid. It is important to establish a uniformly calibrated field for situations when the science goal requires multiple targets in the FOV, or to measure spatial variability in the target.

In the case of an imaging polarimeter, the light is divided over multiple pixels, which is dependent on the sampling and sensitivity of the instrument. As the incoming light beam is split 4 ways, every pixel on the reference channel must be matched to the remaining 3 channels; a pixel-by-pixel ECM will measure how well this matching has been performed, determining how well the system is calibrated. A pixel-by-pixel analysis is the preferred method of analysis when calibrating imaging polarimeters, particularly when there is division of the incoming beam. This is because each pixel can have different properties, such as gain or noise, or the detector chip may contain dead or saturated pixels (no useful information is gathered). Averaging over a number of pixels can help to reduce detector variations, however, this will lead to the loss of polarimetric resolution.

{\scshape pyraf} and {\scshape Matlab} were used to perform the GASP image analysis including bias subtraction, image registration, image combination, calibration, and aperture photometry.  All calibration images were debiased; each detector using its own master bias frame. Flat-fielding is not performed on GASP data as the ECM procedure takes into account any pixel-to-pixel sensitivity variation between images. It does not remove artefacts or optical distortions contributed by the telescope, however, these effects will be present in all channels. {\scshape pyraf} was used to perform image registration using observational target centroid information. The first channel of the reflected path (RP1) was selected as the reference channel, and all other calibration and science channels are mapped to RP1 using {\em geomap}. A systematic error in spatial registration was approximated to be 1 pixel. 

All GASP data was recorded using 2 EMCCD detectors, i.e. both RP channels are recorded on one detector, and the transmitted path (TP) channels on the other. A full description of this reduction can be found in \cite{Kyne14}. Common data points (from science images) were matched between the reference and input images, producing a geometrical transformation which the {\scshape iraf} command {\em gregister} used to register the images. The minimum function that could be used to fit the data was a 3rd order polynomial, which includes higher order distortions in the registration solution. A function of this order requires at least 10 matched points. A data point \lq\lq outlier\rq\rq\ could grossly distort the transformation, therefore RMS value checks were carried out to find the best combination of points. The RMS value of the fit is only significant in the region where the points are located; however, RMS values of matched points that are higher than 1 pixel were normally rejected.

Aperture photometry was carried out on a number of polarization standards. Only two standards were analysed in detail as they were found to have the best observational seeing. A 0\% standard HD12021 and, closer in time to the Crab pulsar observations, a 6\% standard BD25727. Observations of each target were made in both a Clear and a R Band filter. The results of these can be found in Table \ref{table1}.

\begin{table*}
  \centering
 \caption{Polarimetric results for the polarization standards observed by GASP \citep{Kyne14}. The data is sky subtracted, and the polarization angle has been corrected for telescope orientation. %The difference in aperture size taken is a result of a changing FWHM for each target in each filter. 
 A radial aperture equal to the radius of the FWHM was used. The expected result for the polarization standards are also stated \citep{Schmidt_standards,Turnshek_HST_standards}.}
    \begin{tabular}{ccccccccc}
    \hline
    \hline
    Target & Band  &       & Expected &       &       & GASP  &       & Reference \\

          			&       		& P(\%)			&$\rm \theta$ ($^{\circ}$)		& P$\rm_{circ.}$ (\%)		& P(\%)			&$\rm \theta$ ($^{\circ}$)	& P$\rm_{circ.}$ (\%) 			&  \\
   \hline
    HD12021 		& R     		& -     			& -     					& -     				& 7.35$\pm$0.07	& 53.76$\pm$0.71 		& 1.01$\pm$0.15		& [1] \\
          			& V     		& 0.08$\pm$0.02    	&160.10$\pm$6.61   		& -     				& 6.82$\pm$0.20	& 87.84$\pm$0.57		& 5.42$\pm$0.13		& [2,1] \\
         				& B     		& 0.11$\pm$0.03 	&169.24$\pm$6.40 			& -     				& -     			& -					& -					& [2,3] \\
          \hline
    BD25727 		& R     		& 6.39$\pm$0.04  	&32.6$\pm$0.2				& -     				& 6.65$\pm$0.49  	& 29.26$\pm$1.43 		& -0.89$\pm$0.28		& [3,1] \\
          			& V     		& 6.15$\pm$0.09	&32.6$\pm$0.4				& -     				& 7.10$\pm$0.12   	& 33.48$\pm$0.70 		& -0.19$\pm$0.16		& [3,1] \\
    \hline
    \end{tabular}%
  \label{table1}%
     \medskip
      \medskip
\begin{flushleft}
\vspace{-0.75\skip\footins}
Reference: [1]  \citet{Kyne14}, [2] \citet{Schmidt_standards},  [3] \citet{Turnshek_HST_standards}
\end{flushleft}
\end{table*}% 

The size of the radial aperture was chosen based on the percentage of light contained in the FWHM of the target, after fitting a radial profile to the PSF. It was found that the FWHM showed very little variation in size between all 4 channels after calibration. The total flux was calculated from each channel for these targets. A sky background annulus was fitted so that the wings of the targets were not sampled -- an area free from target light. It was found that sky subtraction made little difference to the polarimetric result. It is clear that the results from HD12021 are not comparable with that of \cite{Schmidt_standards} or \cite{Turnshek_HST_standards}.  \cite{Kyne14} shows that analysis of this target is subject to instrumental polarization, as such this is a weakly polarized object observed during cloudy conditions. According to \cite{Patat2006}, the instrumental polarization is not removed by the sky background subtraction. Instrumental polarization is independent of the object's intensity. Based on an instrumental polarization $\rm p$, and an instrumental polarization angle $\rm \varphi$,

\begin{equation}
\rm P = \sqrt{P_{0}^{2} + p^{2} + 2P_{0}p\cos{2(\chi_{0} - \varphi)}},
\label{P_greater}
\end{equation}

where $\rm P$ is the observed polarization, and $\rm P_{0}$ and $\rm \chi_{0}$ are the expected polarization and polarization angle, respectively. From this expression, it is clear that when $\rm P_{0} \gg p$, then $\rm P \simeq P_{0}$. In the case that an object and the measured instrumental polarization are comparable ($\rm p \simeq P_{0}$), the observed polarization is approximately given by,

\begin{equation}
\rm P = \sqrt{2}P_{0}\sqrt{1 + \cos(2(\chi_{0} - \varphi))}.
\label{P_equal}
\end{equation}

The main difference between instrumental polarization and a polarized background is that the latter is effective only when the background is $\rm \gtrsim I$ (\citealt{Patat2006} discuss this in more detail), while the former acts regardless of the object intensity; the important element in the measurement is its polarization. The authors also discuss how an average instrumental polarization can be calculated when a known source of a particular polarization is observed. Therefore, in the case of polarimetric observations with GASP, in poor conditions, weakly polarized targets will most probably be affected in this way. HD12021 was used to measure the instrumental polarization as it is a 0\% polarization standard. Equation \ref{P_equal} is used for $\rm p \simeq P_{0}$ for the clear filter observation. The polarization that should have been measured by GASP is 0.07$\pm$0.2\%. Therefore, it is expected that if HD12021 was observed again, in similar conditions, that it would measure a value for the polarization of 6.82$\pm$6.80\%. A similar approach is applied to BD25727. Using the same filter and Equation \ref{P_equal} as $\rm p \simeq P_{0}$ , i.e. the instrumental polarization is equal to the expected value in the literature, and this gives a polarization of 7.0\%. This is in agreement with what was observed by GASP. The same approach is used for the R Band filter, see \cite{Kyne14} for more detail.

The results of this instrumental polarization check indicate that this calculation is only valid for sources which are weakly polarized where GASP overestimates the polarization. The value for the polarization is biased in such a way that it can never be negative, so any error contributors are added. \cite{Collins13} discusses this using polarimetric simulations for various noise affects. It is also possible that the instrumental polarization changed over the course of the night, but we did not repeat the HD12021 observation. The atmospheric variation between these targets could also lead to an error in using this instrumental polarization.

The flux from each channel, I, is converted to a vector and a system matrix \citep{Kyne14}, A, is used to calculate the Stokes vector, S. The relationship can be found below:

\begin{equation}
\rm S = A^{-1} I \\
\end{equation}

The same approach was used to calculate the flux for the pulsar for each channel. However, sky subtraction was not carried out as it was not possible to find an area of the FOV that was flat enough, and free from light from the surrounding nebula. Light from the wings of the pulsar and Trimble 28 extended too far into the surrounding sky. A standard aperture photometry method was used to calculate the total flux from each channel for the pulsar. 

The optical observations by GASP were subject to errors from optical distortion. This meant that there was a limited FOV where a flat sky field could be used to determine a sky value for subtraction. Hence, the polarimetric results for the Crab pulsar and Trimble 28 were not sky subtracted. The analysis from the GASP polarimetric standards imply that sky subtraction shows minor changes to the polarimetry. It was not possible to find a flat section of the nebula to find consistent polarimetric results, and an area of the FOV that was not contaminated by flux from the wings of the targets. Trimble 28 has been measured by HST/ACS, and was found to show very low polarization -- it could be considered a 0\% standard, particularly as it was found to fluctuate over time. The value for the polarization angle varies greatly (expected for a 0\% standard) and the value found by GASP falls close to this measured range. An instrumental polarization measurement was also applied to this data. The R Band filter data for HD12021 was used as a comparison, and the result from \cite{Slowikowska09} and \cite{Moran13} used as the expected measurement. Using Equation \ref{P_greater} an expected value of 0.1\% for Trimble 28 is calculated. The Crab pulsar was also examined using Equation \ref{P_equal}. A polarization of 10.2\% was calculated, within the calculated error limits. The measurement for Trimble 28 is not conclusive; it has yet to be confirmed what are the lower limits for instrumental effects, however, as it is expected to be weakly polarized it should be treated in the same way as HD12021.

As a comparison to the GASP measurement of the pulsar+knot polarization with pre 2012 observations, we re-examined the ten HST/ACS observations of the Crab from 2005 September 06 to 2005 November 25 inclusive. In order to determine the polarization of the pulsar+knot, aperture photometry was first performed on the pulsar and synchrotron knot in each polarized image (0\degree, 60\degree and 120\degree) using the {\scshape iraf} task {\em phot} with a radius of 0.8$\arcsec$, which fully encapsulates the flux from both the pulsar and knot. These flux values were then converted into the corresponding Stokes parameters and hence degree of linear polarization and polarization position angle \citep{Moran13}. The sky counts were measured in a region close to, but beyond the pulsar and knot. The mean and standard error of these 10 pulsar+knot polarization measurements are quoted in Table \ref{table2}.

{\bf   It is noted that  \citep{Patat2006} does not include an instrumental error calculation for Stokes V. To date, the literature provides little reference to an analytical solution for instrumental circular polarisation error, as it does for linear. There are references to measurements for faint standard stars, white dwarfs, and some quasars, but only measured values are quoted. These circular polarisation values are very low in point sources and some work has been done on quasars, much less that 1\% \citep{Rich73,Impey95,Hutsemekers14}. We do not anticipate that the 1\% level of circular polarisation detected in this work will change the \citep{Patat2006} analysis of the instrumental error in linear polarisation significantly. However more work is required, beyond the scope of this paper, to confirm this.}

\subsection{Integral Data}

The Crab pulsar/nebula is regularly observed by Integral since its launch in 2002. This resulted in the first measurement of the Crab polarization in the gamma-ray band using the satellite's telescopes, the SPectrometer on Integral (SPI) \citep{Dean08} and the Imager on Board the Integral Satellite (IBIS) \citep{Forot08}. 
In the present work, phase-averaged polarization analysis of the system was made using data collected by IBIS during the first four years of operation of Integral (period 2003--2007) and from 2012 March to 2014 October. These periods were chosen to fit with the optical HST and GASP observations. The evolution of the $\gamma$-ray polarization with time, as measured by Integral, will be analyzed in detail in a further paper. The 2003--2007 observations were summed over 1,120,500 s, whereas the 2012--2014 ones were obtained summing 1,831,030 s of data, in order to have a sufficient signal to noise ratio for these two periods. To get the polarization signal, we use the same analysis procedure and software as the ones described in \citet{Forot08}.

\begin{figure}
\centering
\includegraphics[width=85mm]{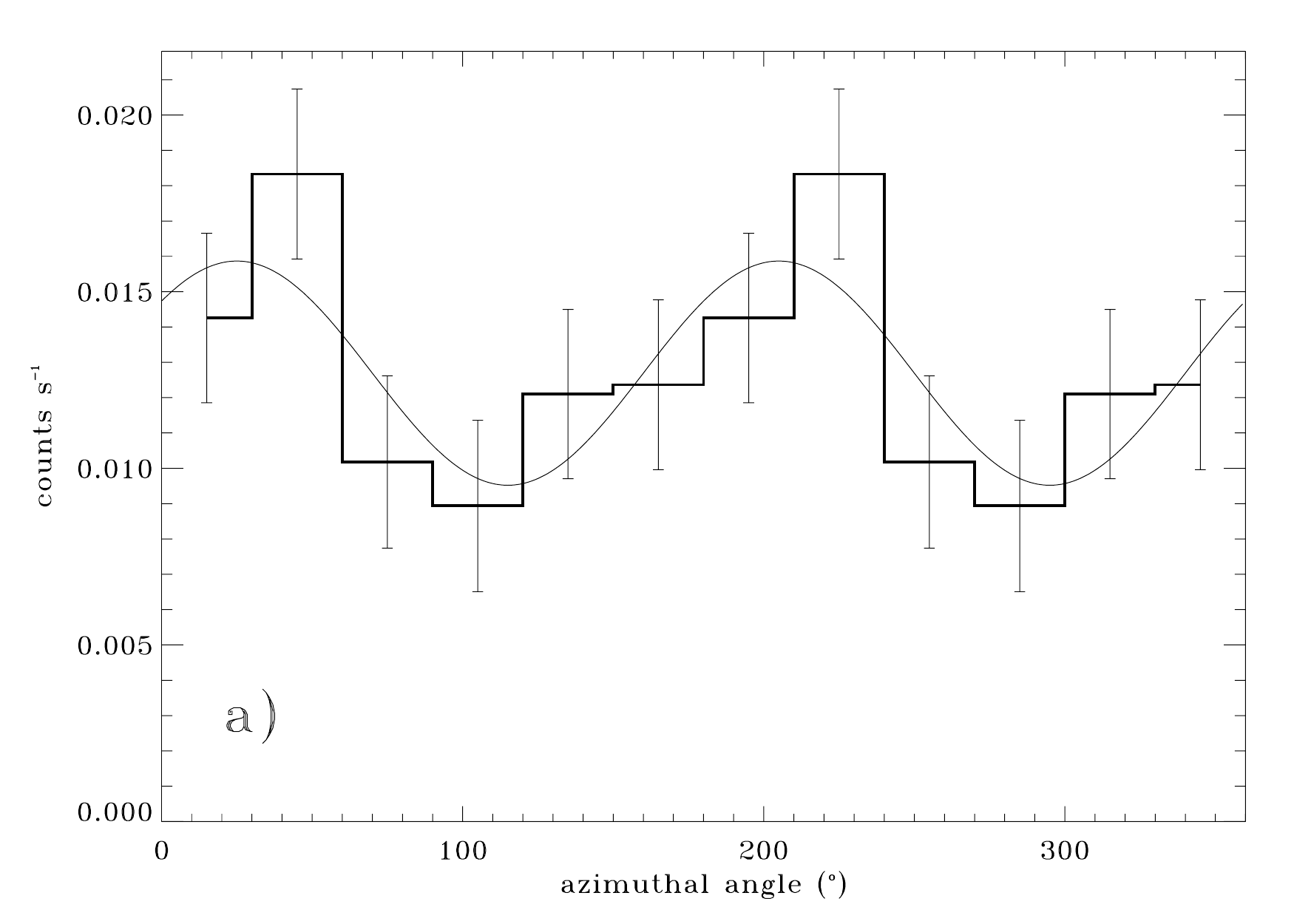}
\includegraphics[width=85mm]{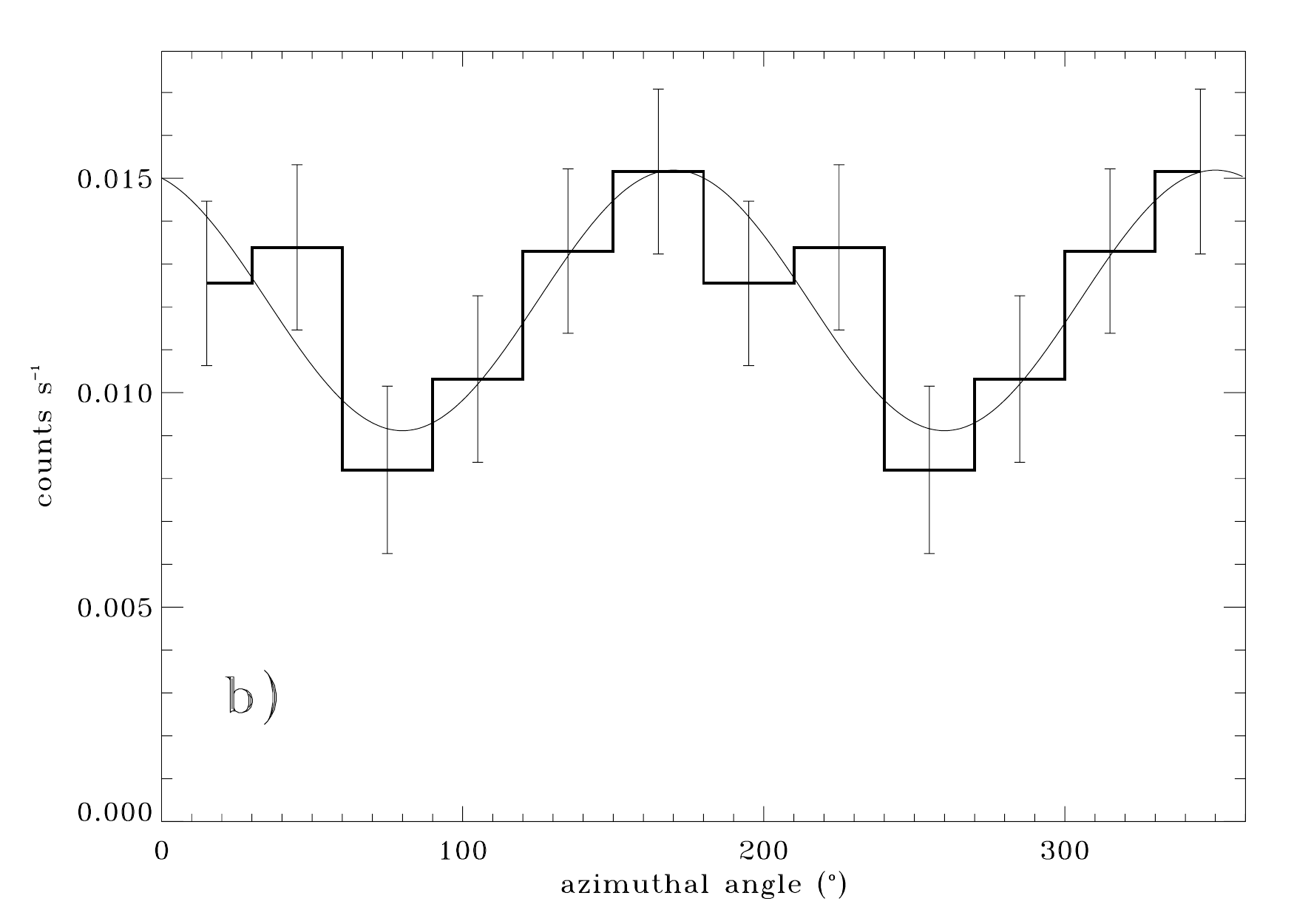}
\caption{Phase-averaged polarization diagram of the Crab pulsar/nebula obtained by Integral/IBIS  at two epochs in the 300--450 keV energy band. The distribution gives the source count rate by azimuthal angle of the Compton scattering. a) phase-averaged signal obtained in the period 2003--2007 b) the same for data collected between 2012 and 2014. The minimum of the sinusoidal fit on both curves, respectively 115\degree and 80\degree, indicates a shift between the two periods of 35\degree ~in the polarization angle.}
\label{Figure4}
\end{figure}

IBIS, in its Compton mode, is a coded-aperture Compton telescope. This design provides high-energy response, low background, and a wide field-of-view. Moreover, utilizing the imaging properties of the coded mask, a high angular resolution for gamma-ray astronomy and background subtraction is achieved. The polarization of celestial sources is measured using Compton scattering properties. Photons entering IBIS are Compton scattered in ISGRI (Integral Soft Gamma-Ray Imager) \citep{Lebrun03}, the first detector plane, at a polar angle $\theta_{com}$, from their incident direction and at an azimuth, $\psi$, from their incident electric vector (see Figure 1 of \citealt{Forot07}). The photons are then absorbed in the second detector, PiCsIT (Pixelated Cesium Iodide Telescope) \citep{Dicocco03}. Hence, the polarization of sources can be measured, since the scattering azimuth is related to the polarization direction. Indeed, the azimuthal profile N$(\psi)$, in Compton counts recorded per azimuth bin, follows  the relation

\begin{equation}
\rm N(\psi) = S [1 + a_{0} cos(2\psi - 2\psi_0)]\\
\end{equation}
for a source polarized at an angle PA = $\psi_{0}$ and with a polarization fraction PF = a$_{0}$/a$_{100}$. The a$_{100}$ amplitude is the measured amplitude for a 100\% polarized source. S is the source mean count rate. The IBIS and SPI polarization capabilities were not calibrated on the ground, due to the tight planning of the satellite. For IBIS, a$\rm_{100}$ has been evaluated to be 
0.29$\pm$0.03 based upon GEANT3 Monte-Carlo simulations of a E$^{-2.2}$ Crab-like spectrum between 300 and 450 keV. To measure N($\psi$), one first needs to optimize the source signal to noise ratio. We thus use only observations at off-axis angles $<$ 7$^{\circ}$ in the 300--450 keV energy range. Based on these selection criteria, the Crab was detected with a signal to noise of 12.9 during the 2003--2007 observations, and of 15.4 during the 2012--2014 period.

After data selection, the ISGRI detector images are decoded to remove the background and get the source counts. The whole process is applied for events in regularly spaced bins in azimuth to derive N($\psi$). The errors on N($\psi$), are dominated by statistic fluctuations in the background dominated observations. Confidence intervals on $a_0$ and $\psi_0$ are not given by the N$(\psi)$ fit to the data since the variables are not independent. They have been derived from the probability density distribution of measuring a and $\psi$ from N$\rm_{pt}$ independent data points in N$(\psi)$  over a $\pi$ period, based on Gaussian distributions for the orthogonal Stokes components \citep{Vinokur65, Vaillancourt06, Forot07,Maier14}:

{ \begin{eqnarray}
&\rm dP(a,\psi) = \frac{N_{pt}~S^2}{\pi~\sigma_S^2} \times \nonumber \\
&\rm exp[-\frac{N_{pt}~S^2}{2~\sigma_S^2}[a^2+a_0^2-2aa_0cos(2\psi-2\psi_0)]]~a~da~d\psi
\end{eqnarray}
\noindent
where   $\rm \sigma_S$ notes the error on the profile mean S. The errors on each a or $\psi$ dimension are obtained by integrating dP(a,$\psi$) over  the other dimension, as it is shown in the paragraph below.

\section{Results}

The results of our analysis are shown in Tables 2 and 3. The optical polarization angle changed from $109.5\pm0.7\degree$ in 2005 (HST/ACS) to $85.3\pm1.4\degree$ in 2012 (GASP observations). The polarization fraction was measured respectively to be 7.7$\pm$0.1\% and 9.6$\pm$0.5\%. Figure \ref{Figure4} shows the polarization diagram for the period 2003--2007 and for 2012--2014 of the Integral $\gamma$-ray observations. The $\gamma$-ray polarization angle changed from 115$\pm$11\degree ~in 2003--2007  with a polarization fraction of 96$\pm$34\% to a position angle of 80$\pm$12\degree ~in 2012--2014 with  nearly the same fraction (98$\pm$37\%). We also observed a non-zero circular optical polarization from the Crab pulsar+knot.

%{\ To examine the validity of the GASP result for the pulsar, we compared the measured optical polarization of the pulsar to the expected source polarization. To calculate this we used equation 2 together with the observed polarization of 9.6$\pm$ 0.5\%, the expected polarization angle of 109$\pm$0.7$^{\circ}$, and the instrumental polarization angle of 53.76$\pm$0.71$^{\circ}$ (HD12021 in the R-band). This yielded an expected polarization of 8.6\%. This is in reasonable agreement with the value measured by using the HST/ACS of 7.7$\pm$0.1\%. We also determined a value for the instrumental polarization angle, $\phi$. Again using equation 2, the observed polarization 9.6$\pm$0.5\%, the expected polarization of 7.7$\pm$0.1\%, and the expected polarization angle of 109$\pm$ 0.7$^{\circ}$ gave a value of 58.6$^{\circ}$. This is comparable to the polarization angle of the unpolarized standard, HD12021, of 53.76$\pm$0.71$^{\circ}$ in the R-band. Lastly, we estimate the expected polarization angle. Again we use equation 2, the observed polarization of 9.6$\pm$0.5\%, the expected polarization of 7.7$\pm$ 0.1\%, and the instrumental polarization angle of 53.76$\pm$0.71$^{\circ}$. This gave a value of 105$^{\circ}$, which deviates from the measured GASP value of 85.3$\pm$1.4$^{\circ}$.}

To check the significance of the shift in the $\gamma$-ray band, we show in Figure \ref{Figure5} the probability density for the two sets of Integral observations together with the 2005 and 2012 optical polarimetric measurements. These plots give the probability density in the azimuthal angle $\psi$, computed by integrating dP in  equation 5 over a, that the real polarization angle is $\psi$ given the data, whatever the polarization fraction might be. As the polarization angle and fraction are not independent variables, these curves are not exactly Gaussian. There is an 11\% probability that the real polarization angle is greater than 100\degree ~for the 2012--2014 observations where a polarization angle of 80\degree ~was measured. In Figure \ref{Figure6}  we show the change in optical and $\gamma$-ray polarization since the original Integral observations \citep{Dean08, Forot08}. Also shown is the change in the optical polarization over the same period. Both wavebands show a similar shift in the angle of polarization angle; $\gamma$-ray 35$\pm$16\degree ~and optical 24.2$\pm$1.6\degree. This is not, within our errors, accompanied by a change in the percentage polarization in either waveband. 

Figure \ref{Figure7} shows the Crab flux above 100 MeV from Fermi and Agile data. One can clearly see the reported periods of flaring. Overlaid on this plot are the polarization position angles of the pulsar+knot (phase-averaged optical and Integral data) (see Tables 2 and 3) during this same period.

\begin{figure}
\centering
\includegraphics[width=85mm]{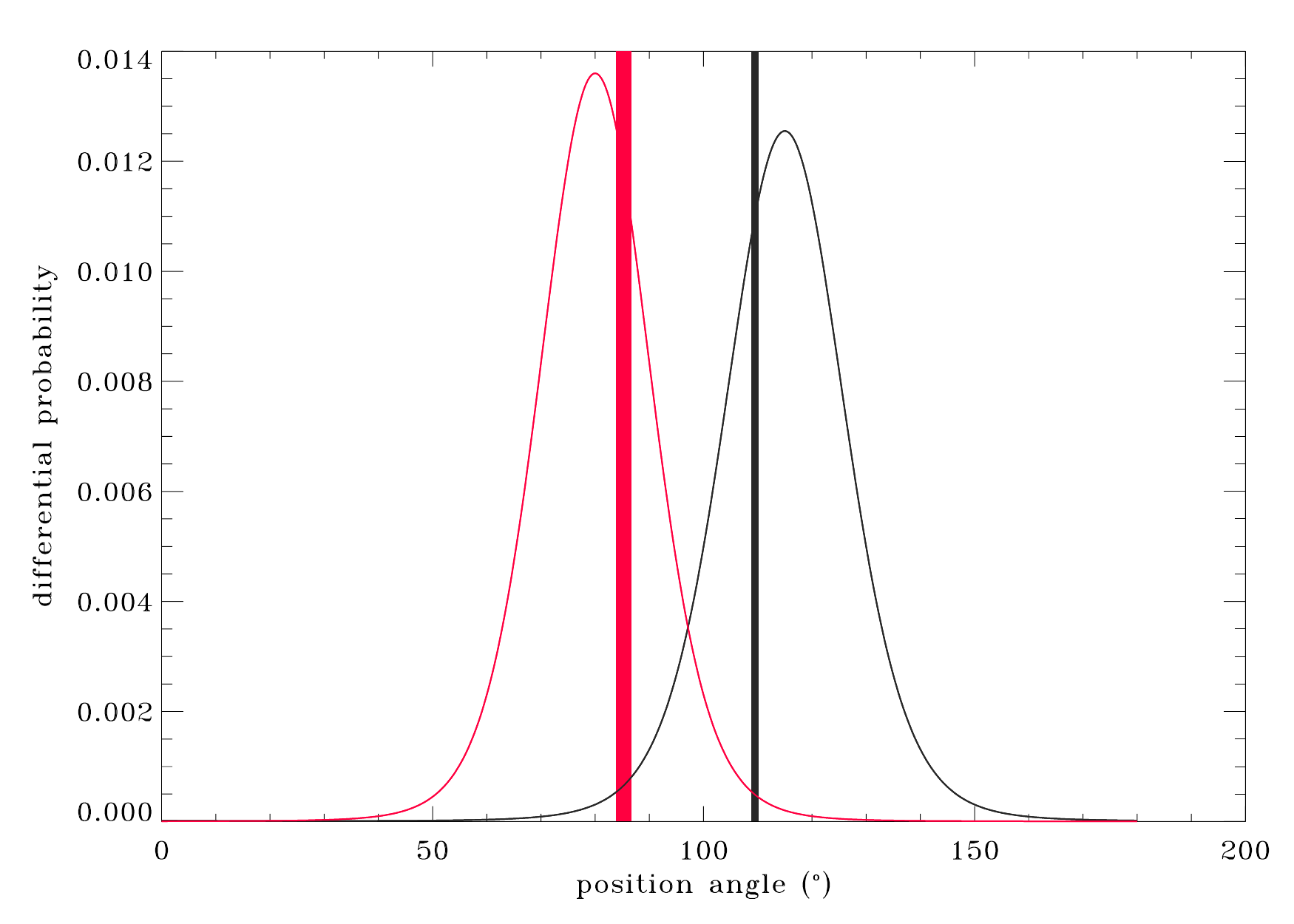}
\caption{The probability density for the two sets of Integral/IBIS observations. The red curve is for the 2012--2014 observations which gives a measured polarization angle of 80\degree ~and the black one for the 2003--2007 set of data with a measured polarization angle  of 115\degree. The vertical bars superimposed on these two distributions are the polarization angle measurements of the HST/ACS 2005 (in black) and GASP 2012 (in red) data (see Table 2). }
\label{Figure5}
\end{figure}

\begin{figure}
\centering
\includegraphics[width=85mm]{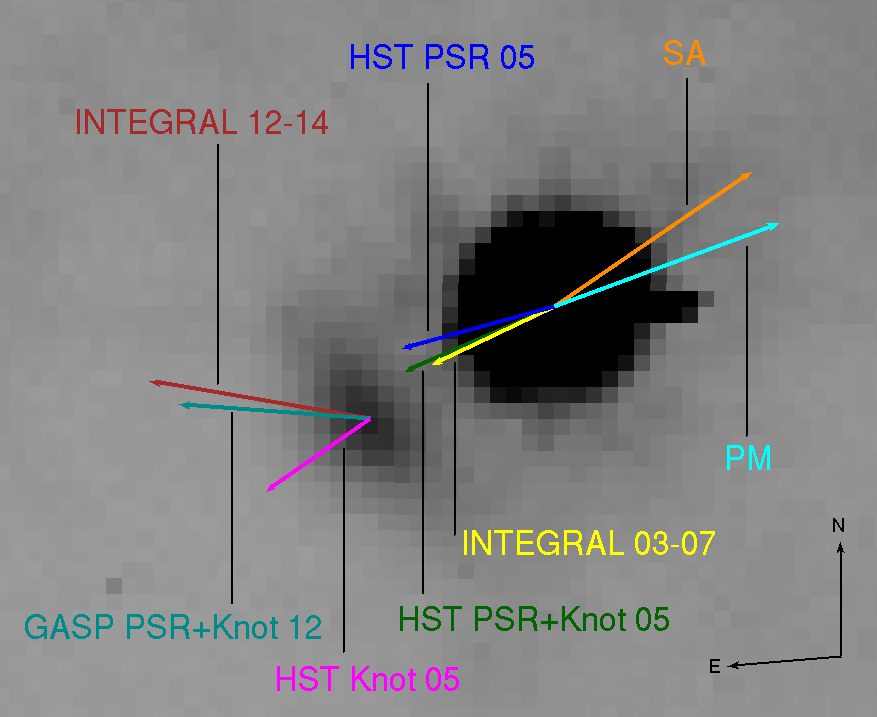}
\caption{Crab polarization vectors taken from Table 2. What is shown are the spin axis (SA) and proper motion (PM) of the pulsar \citep{Kaplan08} with the polarization angles. 
Also of note is that the HST Knot 2005 is similar to the HST PSR+Knot 2005 indicating the pulsar does not contribute significantly to the polarization angle. This is to be expected as the pulsar has 5\% linear polarization over a restricted phase range whilst the knot is polarized at the 60\% level \citep{Slowikowska09,Moran13}. See Figure 5 for clarity on the errors in the polarization position angles. For a wider view of the central Crab nebula see Figure 3.}
\label{Figure6}
\end{figure}

\begin{figure}
\centering
\includegraphics[width=85mm]{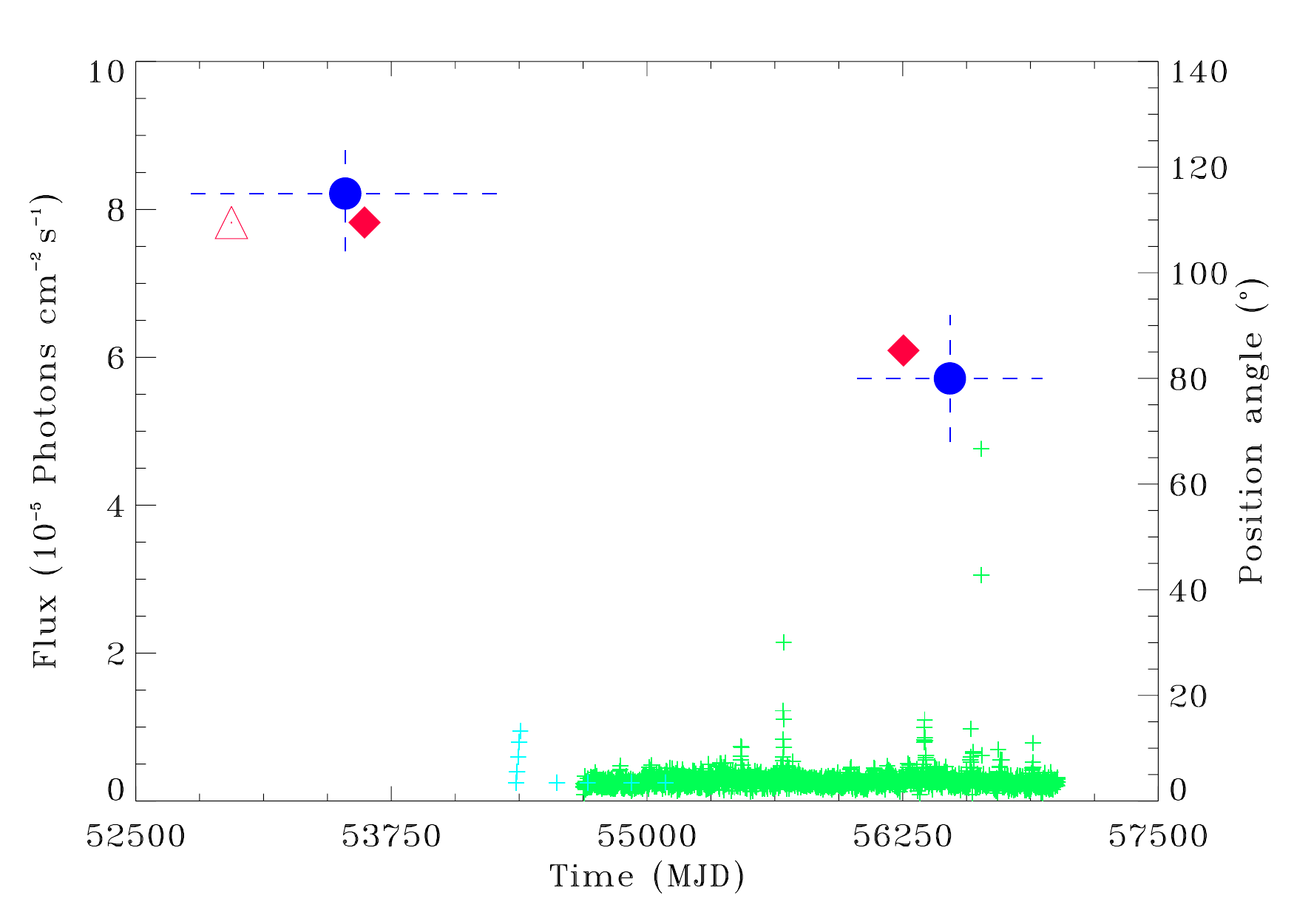}
\caption{
Crab light curve with flux above 100 MeV and polarization measurements obtained between 2003 and 2014 in the optical and Integral bands. The Fermi data are plotted in green and Agile in cyan. The polarization position angles of the pulsar+knot (phase-averaged data) and their respective errors are shown in red for the optical points 
(open triangle from \citet{Slowikowska09}, filled diamond from this work) and in blue filled 
circle for the Integral/IBIS observations (see Tables 2 and 3).
}

\label{Figure7}
\end{figure}

\begin{table*}

\centering
 \begin{minipage}{20cm}
\caption{Summary of the multi-wavelength polarimetry of the Crab nebula and pulsar. }
\begin{tabular}{lcccccc}
\hline
\hline
Waveband &Instrument& Observation&Component &Polarization (\%) & Position Angle (\degree) &Reference\\

& & Year & & & & \\
\hline
$\gamma$-ray & Integral/IBIS & 2003-07 & Phase-avg\footnote{The phase-avg refers to the average polarization over many pulsar rotations} 300-450 keV & 96$\pm$34 & 115$\pm$11 &[1]\\
$\gamma$-ray & Integral/IBIS & 2012-14 & Phase-avg 300-450 keV & 98$\pm$37 & 80$\pm$12 &[1]\\
Optical & RGO-PP/INT & 1988 & Phase-avg & 9.8 & 117 & [2]\\
Optical & HST/ACS & 2005 & Pulsar & 5.2$\pm$0.3 & 105.1$\pm$1.6 & [3]\\ 
Optical & HST/ACS & 2005 & Knot & 59.0$\pm$1.9 & 124.7$\pm$1.0 & [3]\\
Optical & HST/ACS & 2005 & Pulsar+Knot & 7.7$\pm$0.1 & 109.5$\pm$0.7 & [1]\\
Optical & GASP/Hale 200$^{\prime\prime}$ & 2012 & Pulsar+Knot & 9.6$\pm$0.5 & 85.3$\pm$1.4 & [1]\\
Optical & GASP/Hale 200$^{\prime\prime}$ & 2012 & Pulsar+Knot  (circular pol.) &-1.2 $\pm$0.4 & - & [1] \\
%X-ray & OSO-8 & 1976-77 & Nebula & 19.22$\pm$0.92 & 155.79$\pm$1.37 & [4]\\
\hline
\end{tabular}
\label{table2}
\vspace{-0.75\skip\footins}
\end{minipage}
\medskip
\begin{flushleft}
\vspace{-0.75\skip\footins}
Reference: [1]  This work, [2] \citet{Smith88},  [3] \citet{Moran13}%,  [4] \citet{Weisskopf78}
\end{flushleft}
\end{table*}

\begin{table*}
\begin{center}
\centering
 \caption{Optical polarimetry of the Crab pulsar and knot field of view. The F606W filter corresponds to $\lambda=590.70$ nm and $\Delta\lambda=250.00$ nm.}
 \begin{tabular}{lcccccc}
  \hline
  \hline
  Target & Instrument & Observation & Filter & Polarization (\%) & Position Angle (\degree) &Reference\\
  & & Year & & & & \\
  \hline
   Pulsar & HST/ACS & 2005 & F606W & 5.2$\pm$0.3 & 105.1$\pm$1.6 & \citealt{Moran13}\\
   Pulsar & OPTIMA & 2003 & V+R & 5.4$\pm$0.1 & 96.4$\pm$0.2 & \citealt{Slowikowska09}\\
   Pulsar+Knot & HST/ACS  & 2005 & F606W & 7.7$\pm$0.1 & 109.5$\pm$0.7 & This work\\
  Pulsar+Knot & OPTIMA & 2003 & V+R & 9.8$\pm$0.1 & 109.5$\pm$0.2 & \citealt{Slowikowska09}\\
  Pulsar+Knot & GASP/Hale 200$^{\prime\prime}$ & 2012 & R & 9.6$\pm$0.5 & 85.3$\pm$1.4 & This work\\
  Trimble 28 & GASP/Hale 200$^{\prime\prime}$ & 2012 & R & 6.5$\pm$0.3 & 60.9$\pm$2.7 & This work\\
  \hline
 \end{tabular}
  \label{table3}
 \end{center}
\end{table*}

\section{Discussion}

Our results seem to indicate that the Crab optical polarization angle has changed significantly between 2005 and 2012. We have also observed changes in the polarization angle of the $\gamma$-ray flux. From Table \ref{table2} we see that polarization changes are similar for the optical and high-energy radiation. However, there is disagreement between the OPTIMA (2003) and the HST/ACS (2005) observations of the pulsar+knot in the percentage of linear polarization. This is most likely due to problems associated with comparing the aperture in space with ground based observations that are subject to seeing and scattering. We also note that the  $\gamma$-ray polarization is consistent with the observed optical polarization of the knot rather than the pulsar.

In Table \ref{table2} we have included our measurement of the circular polarization in 2012 which is non-zero (-1.2\% at the 3$\sigma$ level). We note that the degree of circular polarization is low possibly contrary to the the prediction of rising circular polarization with magnetic field strength predicted by \citet{Westfold59} and \citet{Legg68}. A reformulation of the \citet{Legg68} calculation indicates that the circular polarization would be about 1.5\% at a magnetic field strength of about 200 G \citep{deBurca15}. 

Understanding the nature of the $\gamma$-ray flares is a significant theoretical challenge. The individual $\gamma$-ray flares were of relatively short duration, significant flux changes were observed on six hour time scales \citep{Buhler14, Cerutti14}, indicating a small region in the nebula of angular size less than 0.1$\arcsec$. There are additional problems with the maximum energy of the observed photons ($\sim$ 400 MeV), which is above the maximum expected energy for synchrotron radiation of 160 MeV \citep{Cerutti14, Guilbert83}. However, this can be reconciled by Doppler boosting \citep{Guilbert83}. From simultaneous Chandra and Keck telescope observations \citep{Weisskopf13} taken during the April 2011 flare there was no significant increase in the X-ray or near IR flux from the pulsar region. Moreover, longer term observations prior to the 2009-2013 flares indicate that the inner knot is variable at IR wavelengths \citep{Sandberg09}. Explanations for the $\gamma$-ray flux increase have been varied with explosive magnetic recombination events being the preferred model \citep{Buhler14, Cerutti14}. This would be in contrast to shock acceleration which is the dominant mechanism in supernova remnants \citep{Aharonian13}. What is also not clear is where in the nebula the flare emission might originate although it is most likely to come from close to the pulsar.

Changes in the angle and degree of polarization of synchrotron radiation are indicative of variations in the geometry of the local magnetic field and plasma. Our results argue for a change in the magnetic field orientation around the region of the synchrotron-emitting knot close to the pulsar. This could be complemented by changes in the local plasma density or energy distribution. However, as the optical emission from the knot does not seem to vary significantly on short time scales we suggest that we are not observing a change in the density of the plasma associated with the optical emission. Such a change of polarization without a change in intensity would be expected if the $\gamma$-ray flares were triggered by magnetic recombination events. However, recent works \citep{Rudy15, Lyutikov15, Yuan15} based on a multi-wavelength study (without polarization measurements) of the inner part of the Crab nebula and MHD simulations conclude that the knot does not appear to be the source of the $\gamma$-ray flares. The change in polarization as observed in the optical and in the low-energy $\gamma$-ray regime by GASP and Integral might indicate a different process, even linked to magnetic reconnection close to the vicinity of the pulsar, but at a different scale/region/epoch than the $\gamma$-ray flares.

%{\bf As the change in polarization angle is associated with the inner knot it may be the source of at least one, and possibly all, of the Crab $\gamma$-ray flares. However, \citet{Yuan15} have reviewed the implications of recent observations of the Crab nebula and claim that the knot does not appear to be the source of the $\gamma$-ray flares. Hence, though the observed change in polarization may be due to magnetic reconnection, this process may not subsequently be responsible for the $\gamma$-ray flares.}

\section{Conclusions}

For this study we observed similar changes in both the optical and $\gamma$-ray polarization. This would imply a change in the overall magnetic field orientation and consequently gives credence to the source being magnetic reconnection. Furthermore, changes in the optical polarization have been observed to be associated with $\gamma$-ray flares from AGN \citep{Abdo10}. Hence, it is possible that these flaring events are dominated by magnetic reconnection phenomena shown in part by the relatively small changes in the optical/near-IR flux compared to the large changes in the polarization angle. More detailed phase-resolved observations would clear up any uncertainty about whether the pulsar or knot is the source of the circular polarization. From \citet{Westfold59} and \citet{deBurca15} the degree of circular polarization is sensitive to the magnetic field strength particularly in the range from 100--10$^{5}$ G. More observations, particularly phase resolved, are needed before a definitive statement on the local magnetic field strength can be made. 

Our current work represents observations of magnetic reconnection on a much larger scale than previously observed \citep{Benz10}.  It is also of a reconnection event with significantly higher particle energies than seen in the Sun or in stellar flares. More routine optical polarimetric observations  of the Crab nebula would address one of the outstanding questions; whether the optical polarization changes during a $\gamma$-ray flare. Ideally there should be polarimetric observations which straddle the time of $\gamma$-ray flares. \\

\section*{Acknowledgments}

The authors are grateful for time allocated for GASP observations at Palomar. Work partly based on observations with INTEGRAL, a European Space Agency (ESA) project with instruments and science data center funded by ESA member states (especially the Principal Investigator countries: Denmark, France, Germany, Italy, Switzerland, and Spain), Czech Republic and Poland, and with the participation of Russia and the United States. The HST/ACS data used in this paper were obtained from MAST. STScI is operated by the Association of Universities for Research in Astronomy, Inc., under NASA contract NAS5-26555. Science Foundation Ireland is acknowledged for its support in the development of GASP under grant number 09/RFP/AST2391. PM thanks the Irish Research Council for support. This work was made possible in part through support of the Ulysses Ireland-France collaborative funding research program.

\bibliographystyle{mn2e}
\bibliography{reference}

\begin{thebibliography}{}

\bibitem[\protect\citeauthoryear{{Abdo}, {Ackermann}, {Ajello}, {Allafort},
  {Baldini}, {Ballet}, {Barbiellini}, {Bastieri} \& {et al.}}{{Abdo}
  et~al.}{2011}]{Abdo11}
{Abdo} A.~A.,  {Ackermann} M.,  {Ajello} M.,  {Allafort} A.,  {Baldini} L.,
  {Ballet} J.,  {Barbiellini} G.,  {Bastieri} D.,    {et al.} 2011, Science,
  331, 739

\bibitem[\protect\citeauthoryear{{Abdo}, {Ackermann}, {Ajello}, {Axelsson},
  {Baldini}, {Ballet}, {Barbiellini}, {Bastieri},  \& {et al.}}{{Abdo}
  et~al.}{2010}]{Abdo10}
{Abdo} A.~A.,  {Ackermann} M.,  {Ajello} M.,  {Axelsson} M.,  {Baldini} L.,
  {Ballet} J.,  {Barbiellini} G.,  {Bastieri} D.,     {et al.} 2010, Nature,
  463, 919

\bibitem[\protect\citeauthoryear{{Aharonian}}{{Aharonian}}{2013}]{Aharonian13}
{Aharonian} F.~A.,  2013, Astroparticle Physics, 43, 71

\bibitem[\protect\citeauthoryear{{Benz} \& {G{\"u}del}}{{Benz} \&
  {G{\"u}del}}{2010}]{Benz10}
{Benz} A.~O.,  {G{\"u}del} M.,  2010, ARA\&A, 48, 241

\bibitem[\protect\citeauthoryear{{Bietenholz}, {Frail} \&
  {Hester}}{{Bietenholz} et~al.}{2001}]{Bietenholz01}
{Bietenholz} M.~F.,  {Frail} D.~A.,    {Hester} J.~J.,  2001, ApJ, 560, 254

\bibitem[\protect\citeauthoryear{{B{\"u}hler} \& {Blandford}}{{B{\"u}hler} \&
  {Blandford}}{2014}]{Buhler14}
{B{\"u}hler} R.,  {Blandford} R.,  2014, Reports on Progress in Physics, 77,
  066901

\bibitem[\protect\citeauthoryear{{Cerutti}, {Werner}, {Uzdensky} \&
  {Begelman}}{{Cerutti} et~al.}{2014}]{Cerutti14}
{Cerutti} B.,  {Werner} G.~R.,  {Uzdensky} D.~A.,    {Begelman} M.~C.,  2014,
  Physics of Plasmas, 21, 056501

\bibitem[\protect\citeauthoryear{{Collins}, {Kyne}, {Lara}, {Redfern},
  {Shearer} \& {Sheehan}}{{Collins} et~al.}{2013}]{Collins13}
{Collins} P.,  {Kyne} G.,  {Lara} D.,  {Redfern} M.,  {Shearer} A.,
  {Sheehan} B.,  2013, Experimental Astronomy, 36, 479

\bibitem[\protect\citeauthoryear{{Compain}, {Poirer} \& {Drevillon}}{{Compain}
  et~al.}{1999}]{Compain99}
{Compain} E.,  {Poirer} S.,    {Drevillon} B.,  1999, Applied Optics, 38, 3490

\bibitem[\protect\citeauthoryear{{Dean}, {Clark}, {Stephen}, {McBride},
  {Bassani}, {Bazzano}, {Bird}, {Hill}, {Shaw} \& {Ubertini}}{{Dean}
  et~al.}{2008}]{Dean08}
{Dean} A.~J.,  {Clark} D.~J.,  {Stephen} J.~B.,  {McBride} V.~A.,  {Bassani}
  L.,  {Bazzano} A.,  {Bird} A.~J.,  {Hill} A.~B.,  {Shaw} S.~E.,    {Ubertini}
  P.,  2008, Science, 321, 1183

\bibitem[\protect\citeauthoryear{{de B{\'u}rca} \& {Shearer}}{{de B{\'u}rca} \&
  {Shearer}}{2015}]{deBurca15}
{de B{\'u}rca} D.,  {Shearer} A.,  2015, MNRAS, 450, 533-540

\bibitem[\protect\citeauthoryear{{Di Cocco}, \& {et al.}}{{Di Cocco}
  et~al.}{2003}]{Dicocco03}
{Di Cocco} G., {et al.},  2003, A\&A, 411, L189

\bibitem[\protect\citeauthoryear{{Duyvenday}}{{Duyvenday}}{1942}]{Duyvenday42}
{Duyvenday} J.~J.~L.,  1942, PASP, 54, 91

\bibitem[\protect\citeauthoryear{{Forot}, {Laurent}, {Grenier}, {Gouiff{\`e}s}
  \& {Lebrun}}{{Forot} et~al.}{2008}]{Forot08}
{Forot} M.,  {Laurent} P.,  {Grenier} I.~A.,  {Gouiff{\`e}s} C.,    {Lebrun}
  F.,  2008, ApJ, 688, L29

\bibitem[\protect\citeauthoryear{{Forot}, {Laurent}, {Lebrun} \&
  {Limousin}}{{Forot} et~al.}{2007}]{Forot07}
{Forot} M.,  {Laurent} P.,  {Lebrun} F.,    {Limousin} O.,  2007, ApJ, 668,
  1259

\bibitem[\protect\citeauthoryear{{Guilbert}, {Fabian} \& {Rees}}{{Guilbert}
  et~al.}{1983}]{Guilbert83}
{Guilbert} P.~W.,  {Fabian} A.~C.,    {Rees} M.~J.,  1983, MNRAS, 205, 593

\bibitem[\protect\citeauthoryear{{Hester}}{{Hester}}{2008}]{Hester08}
{Hester} J.~J.,  2008, ARA\&A, 46, 127

\bibitem[\protect\citeauthoryear{{Hutsem{\'e}kers}, {Braibant}, {Pelgrims} \&
  {Sluse}}{{Hutsem{\'e}kers} et~al.}{2014}]{Hutsemekers14}
{Hutsem{\'e}kers} D.,  {Braibant} L.,  {Pelgrims} V.,    {Sluse} D.,  2014,
  A\&A, 572, A18

\bibitem[\protect\citeauthoryear{{Impey}, {Malkan}, {Webb} \& {Petry}}{{Impey}
  et~al.}{1995}]{Impey95}
{Impey} C.~D.,  {Malkan} M.~A.,  {Webb} W.,    {Petry} C.~E.,  1995, ApJ, 440,
  80

\bibitem[\protect\citeauthoryear{{Kaplan}, {Chatterjee}, {Gaensler} \&
  {ANDERSON}}{{Kaplan} et~al.}{2008}]{Kaplan08}
{Kaplan} D.~L.,  {Chatterjee} S.,  {Gaensler} B.~M.,    {ANDERSON} J.,  2008,
  ApJ, 677, 1201

\bibitem[\protect\citeauthoryear{{Kirsch}, {Briel}, {Burrows}, {Campana},
  {Cusumano}, {Ebisawa}, {Freyberg}, {Guainazzi} \& {et al.}}{{Kirsch}
  et~al.}{2005}]{Kirsch05}
{Kirsch} M.~G.,  {Briel} U.~G.,  {Burrows} D.,  {Campana} S.,  {Cusumano} G.,
  {Ebisawa} K.,  {Freyberg} M.~J.,  {Guainazzi} M.,    {et al.} 2005, in
  {Siegmund} O.~H.~W.,  ed., UV, X-Ray, and Gamma-Ray Space Instrumentation for
  Astronomy XIV Vol.~5898 of Society of Photo-Optical Instrumentation ENGINEERS
  (SPIE) Conference Series, {Crab: the standard x-ray candle with all (modern)
  x-ray satellites}.
pp 22--33

\bibitem[\protect\citeauthoryear{{Kyne}}{{Kyne}}{2014}]{Kyne14}
{Kyne} G., 2014, PhD Thesis NUI Galway,\footnote{https://aran.library.nuigalway.ie/xmlui/handle/10379/4572}

\bibitem[\protect\citeauthoryear{{Lebrun}, \& {et al.}}{{Lebrun}
  et~al.}{2003}]{Lebrun03}
{Lebrun} F., {et al.},  2011, A\&A, 411, L141

\bibitem[\protect\citeauthoryear{{Legg} \& {Westfold}}{{Legg} \&
  {Westfold}}{1968}]{Legg68}
{Legg} M.~P.~C.,  {Westfold} K.~C.,  1968, ApJ, 154, 499

\bibitem[\protect\citeauthoryear{{Lobanov}, {Horns} \& {Muxlow}}{{Lobanov}
  et~al.}{2011}]{Lobanov11}
{Lobanov} A.~P.,  {Horns} D.,    {Muxlow} T.~W.~B.,  2011, A\&A, 533,
A10

\bibitem[\protect\citeauthoryear{{Lyutikov}, {Komissarov} \&
  {Porth}}{{Lyutikov} et~al.}{2015}]{Lyutikov15}
{Lyutikov} M.,  {Komissarov} S.,    {Porth} O.,  2015, ArXiv e-prints

\bibitem[\protect\citeauthoryear{{Maier}, {Tenzer} \& {Santangelo}}{{Maier} 
et~al.}{2014}]{Maier14}
{Maier} D., {Tenzer} C., {Santangelo} A., 2014, PASP, 126, 459

\bibitem[\protect\citeauthoryear{{Mayall} \& {Oort}}{{Mayall} \&
  {Oort}}{1942}]{Mayall42}
{Mayall} N.~U.,  {Oort} J.~H.,  1942, PASP, 54, 95

\bibitem[\protect\citeauthoryear{{Moran}, {Shearer}, {Mignani},
  {S{\l}owikowska}, {De Luca}, {Gouiff{\`e}s} \& {Laurent}}{{Moran}
  et~al.}{2013}]{Moran13}
{Moran} P.,  {Shearer} A.,  {Mignani} R.~P.,  {S{\l}owikowska} A.,  {De Luca}
  A.,  {Gouiff{\`e}s} C.,    {Laurent} P.,  2013, MNRAS, 433, 2564

\bibitem[\protect\citeauthoryear{{Morii}, {Kawai}, {Usui}, {Sugimori},
  {Sugizaki}, {Mihara}, {Yamamoto}, {Matsuoka} \& {MAXI Team}}{{Morii}
  et~al.}{2011a,b}]{Morii11ab}
{Morii} M.,  {Kawai} N.,  {Usui} R.,  {Sugimori} K.,  {Sugizaki} M.,  {Mihara}
  T.,  {Yamamoto} T.,  {Matsuoka} M., {MAXI Team}, 2011a, Journal of Physics
  Conference Series, 302, 012062

\bibitem[\protect\citeauthoryear{{Morii}, {Kawai}, {Usui}, {Sugimori},
  {Sugizaki}, {Serino}, {Yamamoto}, {Matsuoka} \& {et al.}}{{Morii}
  et~al.}{2011b}]{Morii11b}
{Morii} M., {et al.}, 2011b, PASJ, 63, 821


\bibitem[\protect\citeauthoryear{{Patat} \& {Romaniello}}{{Patat} \&
  {Romaniello}}{2006}]{Patat2006}
{Patat} F.,  {Romaniello} M.,  2006, {Publications of the Astronomical Society
  of the Pacific}, 118, 146

\bibitem[\protect\citeauthoryear{{Rudy}, {Horns}, {DeLuca}, {Kolodziejczak},
  {Tennant} \& {et al.}}{{Rudy} et~al.}{2015}]{Rudy15}
{Rudy} A.,  {Horns} D.,  {DeLuca} A.,  {Kolodziejczak} J.,  {Tennant} A.,
  {et al.} 2015, ArXiv e-prints

\bibitem[\protect\citeauthoryear{{Rich} \& {Williams}}{{Rich} \&
  {Williams}}{1973}]{Rich73}
{Rich} A.,  {Williams} W.~L.,  1973, ApJ, 180, L123

\bibitem[\protect\citeauthoryear{{Sandberg} \& {Sollerman}}{{Sandberg} \&
  {Sollerman}}{2009}]{Sandberg09}
{Sandberg} A.,  {Sollerman} J.,  2009, A\&A, 504, 525

\bibitem[\protect\citeauthoryear{{Scargle}}{{Scargle}}{1969}]{Scargle69}
{Scargle} J.~D.,  1969, ApJ, 156, 401

\bibitem[\protect\citeauthoryear{{Schmidt}, {Elston} \& {Lupie}}{{Schmidt}
  et~al.}{1992}]{Schmidt_standards}
{Schmidt} G.~D.,  {Elston} R.,    {Lupie} O.~L.,  1992, ApJ, 104, 1563

\bibitem[\protect\citeauthoryear{{S{\l}owikowska}, {Kanbach}, {Kramer} \&
  {Stefanescu}}{{S{\l}owikowska} et~al.}{2009}]{Slowikowska09}
{S{\l}owikowska} A.,  {Kanbach} G.,  {Kramer} M.,    {Stefanescu} A.,  2009,
  MNRAS, 397, 103

\bibitem[\protect\citeauthoryear{{Smith}, {Jones}, {Dick} \& {Pike}}{{Smith}
  et~al.}{1988}]{Smith88}
{Smith} F.~G.,  {Jones} D.~H.~P.,  {Dick} J.~S.~B.,    {Pike} C.~D.,  1988,
  MNRAS, 233, 305

\bibitem[\protect\citeauthoryear{{Striani}, {Tavani}, {Vittorini},
  {Donnarumma}, {Giuliani}, {Pucella}, {Argan}, {Bulgarelli} \& {et
  al.}}{{Striani} et~al.}{2013}]{Striani13}
{Striani} E.,  {Tavani} M.,  {Vittorini} V.,  {Donnarumma} I.,  {Giuliani} A.,
  {Pucella} G.,  {Argan} A.,  {Bulgarelli} A.,    {et al.} 2013, ApJ, 765, 52

\bibitem[\protect\citeauthoryear{{Tavani}, {Bulgarelli}, {Vittorini},
  {Pellizzoni}, {Striani}, {Caraveo}, {Weisskopf}, {Tennant} \& {et
  al.}}{{Tavani} et~al.}{2011}]{Tavani11}
{Tavani} M.,  {Bulgarelli} A.,  {Vittorini} V.,  {Pellizzoni} A.,  {Striani}
  E.,  {Caraveo} P.,  {Weisskopf} M.~C.,  {Tennant} A.,    {et al.} 2011,
  Science, 331, 736

\bibitem[\protect\citeauthoryear{{Trimble}}{{Trimble}}{1968}]{Trimble68}
{Trimble} V.,  1968, AJ, 73, 535

\bibitem[\protect\citeauthoryear{{Turnshek}, {Bohlin}, {Williamson} II,
  {Lupie}, {Koornneef} \& {Morgan}}{{Turnshek}
  et~al.}{1990}]{Turnshek_HST_standards}
{Turnshek} D.~A.,  {Bohlin} R.~C.,  {Williamson} II R.~L.,  {Lupie} O.~L.,
  {Koornneef} J.,    {Morgan} D.~H.,  1990, ApJ, 99, 1243

\bibitem[\protect\citeauthoryear{{Ubertini}, {Lebrun}, {Di Cocco}, {Bazzano},
  {Bird}, {Broenstad}, {Goldwurm}, {La Rosa} \& {et al.}}{{Ubertini}
  et~al.}{2003}]{Ubertini03}{Ubertini} P.,  {Lebrun} F.,  {Di Cocco} G.,  {Bazzano} A.,  {Bird} A.~J.,
  {Broenstad} K.,  {Goldwurm} A.,  {La Rosa} G.,    {et al.} 2003, A\&AP, 411,
  L131

\bibitem[\protect\citeauthoryear{{Vedrenne},{Roques},{Sch{\"o}nfelder},{Mandrou}, {Lichti}, {von Kienlin}, {Cordier}, {Schanne} \& {et~al.}}{{Vedrenne} et~al.}{2003}]{Vedrenne03}
{Vedrenne} G., {et al.}, 2003, A\&A, L63

\bibitem[\protect\citeauthoryear{{Vaillancourt}}{{Vaillancourt}}{2006}]{Vaillancourt06}
{Vaillancourt} J. E.,  2006, PASP, 118, 1340

\bibitem[\protect\citeauthoryear{{Vinokur}}{{Vinokur}}{1965}]{Vinokur65}
{Vinokur} M.,  1965, Ann. d'Astrophys., 28, 412


\bibitem[\protect\citeauthoryear{{Weisskopf}, {Guainazzi}, {Jahoda},
  {Shaposhnikov}, {O'Dell}, {Zavlin}, {Wilson-Hodge} \& {Elsner}}{{Weisskopf}
  et~al.}{2010}]{Weisskopf10}
{Weisskopf} M.~C.,  {Guainazzi} M.,  {Jahoda} K.,  {Shaposhnikov} N.,  {O'Dell}
  S.~L.,  {Zavlin} V.~E.,  {Wilson-Hodge} C.,    {Elsner} R.~F.,  2010, ApJ,
  713, 912

\bibitem[\protect\citeauthoryear{{Weisskopf}, {Silver}, {Kestenbaum}, {Long} \&
  {Novick}}{{Weisskopf} et~al.}{1978}]{Weisskopf78}
{Weisskopf} M.~C.,  {Silver} E.~H.,  {Kestenbaum} H.~L.,  {Long} K.~S.,
  {Novick} R.,  1978, ApJL, 220, L117

\bibitem[\protect\citeauthoryear{{Weisskopf}, {Tennant}, {Arons}, {Blandford},
  {Buehler}, {Caraveo}, {Cheung}, {Costa} \& {et al.}}{{Weisskopf}
  et~al.}{2013}]{Weisskopf13}
{Weisskopf} M.~C.,  {Tennant} A.~F.,  {Arons} J.,  {Blandford} R.,  {Buehler}
  R.,  {Caraveo} P.,  {Cheung} C.~C.,  {Costa} E.,    {et al.} 2013, ApJ, 765,
  56

\bibitem[\protect\citeauthoryear{{Westfold}}{{Westfold}}{1959}]{Westfold59}
{Westfold} K.~C.,  1959, ApJ, 130, 241

\bibitem[\protect\citeauthoryear{{Yuan} \& {Blandford}}{{Yuan} \&
  {Blandford}}{2015}]{Yuan15}
{Yuan} Y.,  {Blandford} R.,  2015, ArXiv e-prints

\end{thebibliography}

\label{lastpage}

\end{document}